\pdfoutput=1
\documentclass{moriond}
\usepackage{etoolbox}
\usepackage{eurosym}
\usepackage[export]{adjustbox}
\usepackage{ifthen}
\usepackage{lineno}
\usepackage{mciteplus}
\usepackage{upgreek}
\usepackage{verbatim}
\usepackage{xspace}
\usepackage{subcaption}
\graphicspath{{./figs/}}
\def\lhcb {\mbox{LHCb}\xspace}

\def\lhc    {\mbox{LHC}\xspace}

 \def\Pmu         {\ensuremath{\mu}\xspace}                 
 \def\Pnu         {\ensuremath{\nu}\xspace}                 
                  
 \def\Ppi         {\ensuremath{\pi}\xspace}

 \def\Ptau        {\ensuremath{\tau}\xspace}

 \def\Ppsi        {\ensuremath{\psi}\xspace}                 
                  
 \mathchardef\PDelta="7101
 \mathchardef\PXi="7104
 \mathchardef\PLambda="7103
 \mathchardef\PSigma="7106
 \mathchardef\POmega="710A
 \mathchardef\PUpsilon="7107
                  
 \def\PB      {\ensuremath{B}\xspace}                 
                  
 \def\PD      {\ensuremath{D}\xspace}

 \def\PJ      {\ensuremath{J}\xspace}                 
 \def\PK      {\ensuremath{K}\xspace}

 \def\Pb      {\ensuremath{b}\xspace}                 
 \def\Pc      {\ensuremath{c}\xspace}                 
 \def\Pd      {\ensuremath{d}\xspace}                 
 \def\Pe      {\ensuremath{e}\xspace}

 \def\Pi      {\ensuremath{i}\xspace}

 \def\Pp      {\ensuremath{p}\xspace}

 \def\Ps      {\ensuremath{s}\xspace}

\makeatletter
\ifcase \@ptsize \relax% 10pt
  \newcommand{\miniscule}{\@setfontsize\miniscule{4}{5}}% \tiny: 5/6
\or% 11pt
  \newcommand{\miniscule}{\@setfontsize\miniscule{5}{6}}% \tiny: 6/7
\or% 12pt
  \newcommand{\miniscule}{\@setfontsize\miniscule{5}{6}}% \tiny: 6/7
\fi
\makeatother

\DeclareRobustCommand{\optbar}[1]{\shortstack{{\miniscule (\rule[.5ex]{1.25em}{.18mm})}
  \\ [-.7ex] $#1$}}

\def\epem       {{\ensuremath{\Pe^+\Pe^-}}\xspace}

\def\mun        {{\ensuremath{\Pmu^-}}\xspace} % muon negative (\mum is taken)
\def\mumu       {{\ensuremath{\Pmu^+\Pmu^-}}\xspace}

\def\taum       {{\ensuremath{\Ptau^-}}\xspace}

\def\ellm       {{\ensuremath{\ell^-}}\xspace}
\def\ellp       {{\ensuremath{\ell^+}}\xspace}

\def\neub       {{\ensuremath{\overline{\Pnu}}}\xspace}

\def\neumb      {{\ensuremath{\neub_\mu}}\xspace}

\def\neutb      {{\ensuremath{\neub_\tau}}\xspace}
\def\dquark    {{\ensuremath{\Pd}}\xspace}

\def\squark    {{\ensuremath{\Ps}}\xspace}

\def\cquark    {{\ensuremath{\Pc}}\xspace}

\def\bquark    {{\ensuremath{\Pb}}\xspace}

\def\pion   {{\ensuremath{\Ppi}}\xspace}

\def\pip    {{\ensuremath{\pion^+}}\xspace}

\def\kaon    {{\ensuremath{\PK}}\xspace}
  \def\Kbar    {{\kern 0.2em\overline{\kern -0.2em \PK}{}}\xspace}

\def\KorKbar    {\kern 0.18em\optbar{\kern -0.18em K}{}\xspace}

\def\Kp      {{\ensuremath{\kaon^+}}\xspace}

\def\Kstarz  {{\ensuremath{\kaon^{*0}}}\xspace}

  \def\Dbar    {{\kern 0.2em\overline{\kern -0.2em \PD}{}}\xspace}
\def\D       {{\ensuremath{\PD}}\xspace}

\def\DorDbar    {\kern 0.18em\optbar{\kern -0.18em D}{}\xspace}

\def\Dstar   {{\ensuremath{\D^*}}\xspace}

\def\Dstarp  {{\ensuremath{\D^{*+}}}\xspace}

\def\B       {{\ensuremath{\PB}}\xspace}
\def\Bbar    {{\ensuremath{\kern 0.18em\overline{\kern -0.18em \PB}{}}}\xspace}

\def\BorBbar    {\kern 0.18em\optbar{\kern -0.18em B}{}\xspace}
\def\Bz      {{\ensuremath{\B^0}}\xspace}

\def\Bu      {{\ensuremath{\B^+}}\xspace}

\def\Bd      {{\ensuremath{\B^0}}\xspace}
\def\Bs      {{\ensuremath{\B^0_\squark}}\xspace}

\def\Bdb     {{\ensuremath{\Bbar{}^0}}\xspace}

\def\jpsi     {{\ensuremath{{\PJ\mskip -3mu/\mskip -2mu\Ppsi\mskip 2mu}}}\xspace}
\def\psitwos  {{\ensuremath{\Ppsi{(2S)}}}\xspace}

  \def\Y#1S{\ensuremath{\PUpsilon{(#1S)}}\xspace}% no space before {...}!

\def\proton      {{\ensuremath{\Pp}}\xspace}

\def\Lz          {{\ensuremath{\PLambda}}\xspace}
\def\Lbar        {{\ensuremath{\kern 0.1em\overline{\kern -0.1em\PLambda}}}\xspace}
\def\LorLbar    {\kern 0.18em\optbar{\kern -0.18em \PLambda}{}\xspace}

\def\Lb      {{\ensuremath{\Lz^0_\bquark}}\xspace}

\def\BF         {{\ensuremath{\cal B}}\xspace}

\newcommand{\decay}[2]{\ensuremath{#1\!\to #2}\xspace}         % {\Pa}{\Pb \Pc}

\def\to                 {\ensuremath{\rightarrow}\xspace}

\def\qsq       {{\ensuremath{q^2}}\xspace}

\def\BdToKstmm    {\decay{\Bd}{\Kstarz\mup\mun}}

\def\AT#1     {\ensuremath{A_{\mathrm{T}}^{#1}}\xspace}           % 2

\def\C#1      {\ensuremath{\mathcal{C}_{#1}}\xspace}                       % 9
\def\Cp#1     {\ensuremath{\mathcal{C}_{#1}^{'}}\xspace}                    % 7
\def\Ceff#1   {\ensuremath{\mathcal{C}_{#1}^{\mathrm{(eff)}}}\xspace}        % 9  
\def\Cpeff#1  {\ensuremath{\mathcal{C}_{#1}^{'\mathrm{(eff)}}}\xspace}       % 7
\def\Ope#1    {\ensuremath{\mathcal{O}_{#1}}\xspace}                       % 2
\def\Opep#1   {\ensuremath{\mathcal{O}_{#1}^{'}}\xspace}                    % 7

\newcommand{\tev}{\ensuremath{\mathrm{\,Te\kern -0.1em V}}\xspace}
\newcommand{\gev}{\ensuremath{\mathrm{\,Ge\kern -0.1em V}}\xspace}
\newcommand{\mev}{\ensuremath{\mathrm{\,Me\kern -0.1em V}}\xspace}
\newcommand{\kev}{\ensuremath{\mathrm{\,ke\kern -0.1em V}}\xspace}
\newcommand{\ev}{\ensuremath{\mathrm{\,e\kern -0.1em V}}\xspace}
\newcommand{\gevc}{\ensuremath{{\mathrm{\,Ge\kern -0.1em V\!/}c}}\xspace}
\newcommand{\mevc}{\ensuremath{{\mathrm{\,Me\kern -0.1em V\!/}c}}\xspace}
\newcommand{\gevcc}{\ensuremath{{\mathrm{\,Ge\kern -0.1em V\!/}c^2}}\xspace}
\newcommand{\gevgevcccc}{\ensuremath{{\mathrm{\,Ge\kern -0.1em V^2\!/}c^4}}\xspace}
\newcommand{\mevcc}{\ensuremath{{\mathrm{\,Me\kern -0.1em V\!/}c^2}}\xspace}

\def\invfb   {\ensuremath{\mbox{\,fb}^{-1}}\xspace}

\newcommand{\stat}{\ensuremath{\mathrm{\,(stat)}}\xspace}
\newcommand{\syst}{\ensuremath{\mathrm{\,(syst)}}\xspace}

\def\gsim{{~\raise.15em\hbox{$>$}\kern-.85em
          \lower.35em\hbox{$\sim$}~}\xspace}
\def\lsim{{~\raise.15em\hbox{$<$}\kern-.85em
          \lower.35em\hbox{$\sim$}~}\xspace}

\def\tell1  {TELL1\xspace}
\def\ukl1   {UKL1\xspace}

\def\ll{\ensuremath{\ellp\ellm}\xspace}

\def\bTosll{\decay{\bquark}{\squark \, \ll}}
\def\bTodll{\decay{\bquark}{\dquark \, \ll}}

\def\pp{\ensuremath{\proton\proton}\xspace}

\def\RK{\ensuremath{R_{\kaon}}\xspace}
\def\RKst{\ensuremath{R_{\Kstarz}}\xspace}
\def\RPhi{\ensuremath{R_{\phi}}\xspace}
\def\RDst{\ensuremath{R_{\Dstar}}\xspace}

\def\afb{\ensuremath{A_{FB}}\xspace}
\def\fl{\ensuremath{F_{L}}\xspace}
\def\pfp{\ensuremath{P_{5}'}\xspace}

\def\BdToKstll{\decay{\Bd}{\Kstarz \ll}}
\def\BdToKstmm{\decay{\Bd}{\Kstarz \mumu}}
\def\BdToKstee{\decay{\Bd}{\Kstarz \epem}}

\def\BdToKstJPsll{\decay{\Bd}{\Kstarz \jpsi(\decay{}{\ll})}}
\def\BdToKstJPsmm{\decay{\Bd}{\Kstarz \jpsi(\decay{}{\mumu})}}
\def\BdToKstJPsee{\decay{\Bd}{\Kstarz \jpsi(\decay{}{\epem})}}

\def\BuToKmm{\decay{\Bu}{\Kp \mumu}}
\def\BuToKee{\decay{\Bu}{\Kp \epem}}
\def\BuToPimm{\decay{\Bu}{\pip \mumu}}

\def\BsToPhimm{\decay{\Bs}{\phi \mumu}}

\def\LbToLmm{\decay{\Lb}{\Lz \mumu}}

\newcommand{\Photo}{}

\begin{document}

\vspace*{4cm}
\title{Status of New Physics searches with \bTosll transitions @ LHCb}

\author{Simone Bifani \\ On behalf of the LHCb collaboration.}
\address{University of Birmingham, School of Physics and Astronomy, Birmingham, United Kingdom}

\maketitle\abstracts{Rare decays of heavy-flavoured particles provide an ideal laboratory to look for deviations from the Standard Model, and explore energy regimes beyond the \lhc reach.
Decays proceeding via electroweak penguin diagrams are excellent probes to search for New Physics, and \mbox{\bTosll} processes are particularly interesting since they give access to many observables such as branching fractions, asymmetries and angular observables.
Recent results from the \lhcb experiment are reviewed.}
%%%%%%%%%%%%%%%%%%%%%%%%%%%%%%%%%%%%
\section{Introduction}

\begin{figure}[h!]
\vspace{0.5cm}
\centering
\hspace{-2cm}
\begin{subfigure}[b]{.45\textwidth}
\centering
\includegraphics[width=.5\textwidth]{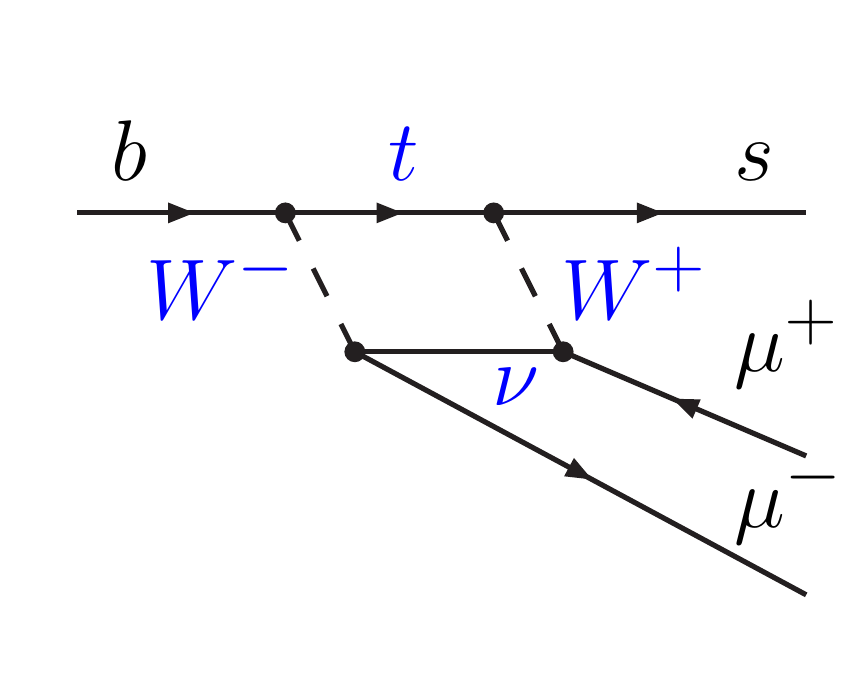}
\vspace{1ex}
\includegraphics[width=.5\textwidth]{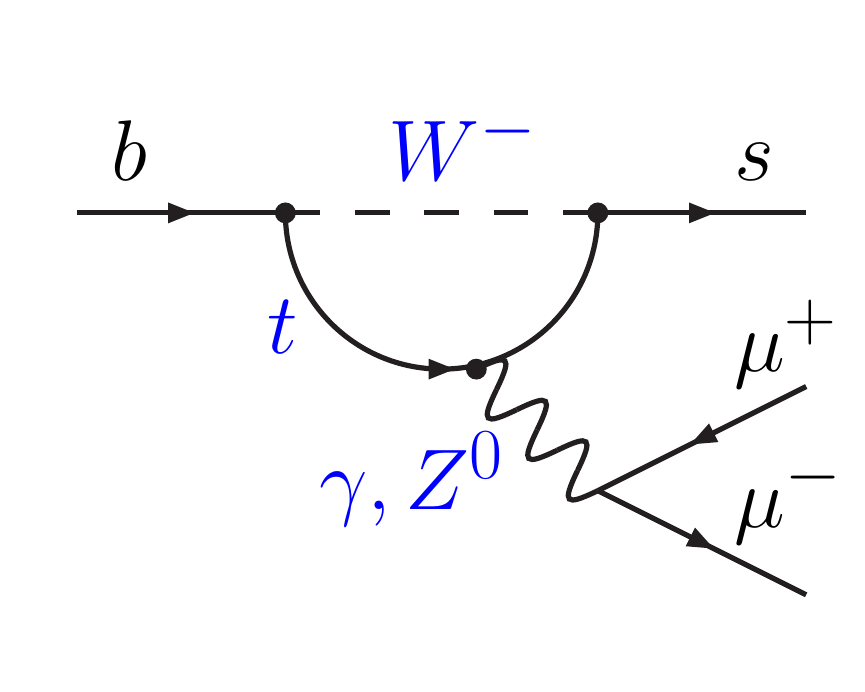}
\end{subfigure}
\begin{subfigure}[b]{.45\textwidth}
\centering
\includegraphics[width=\textwidth]{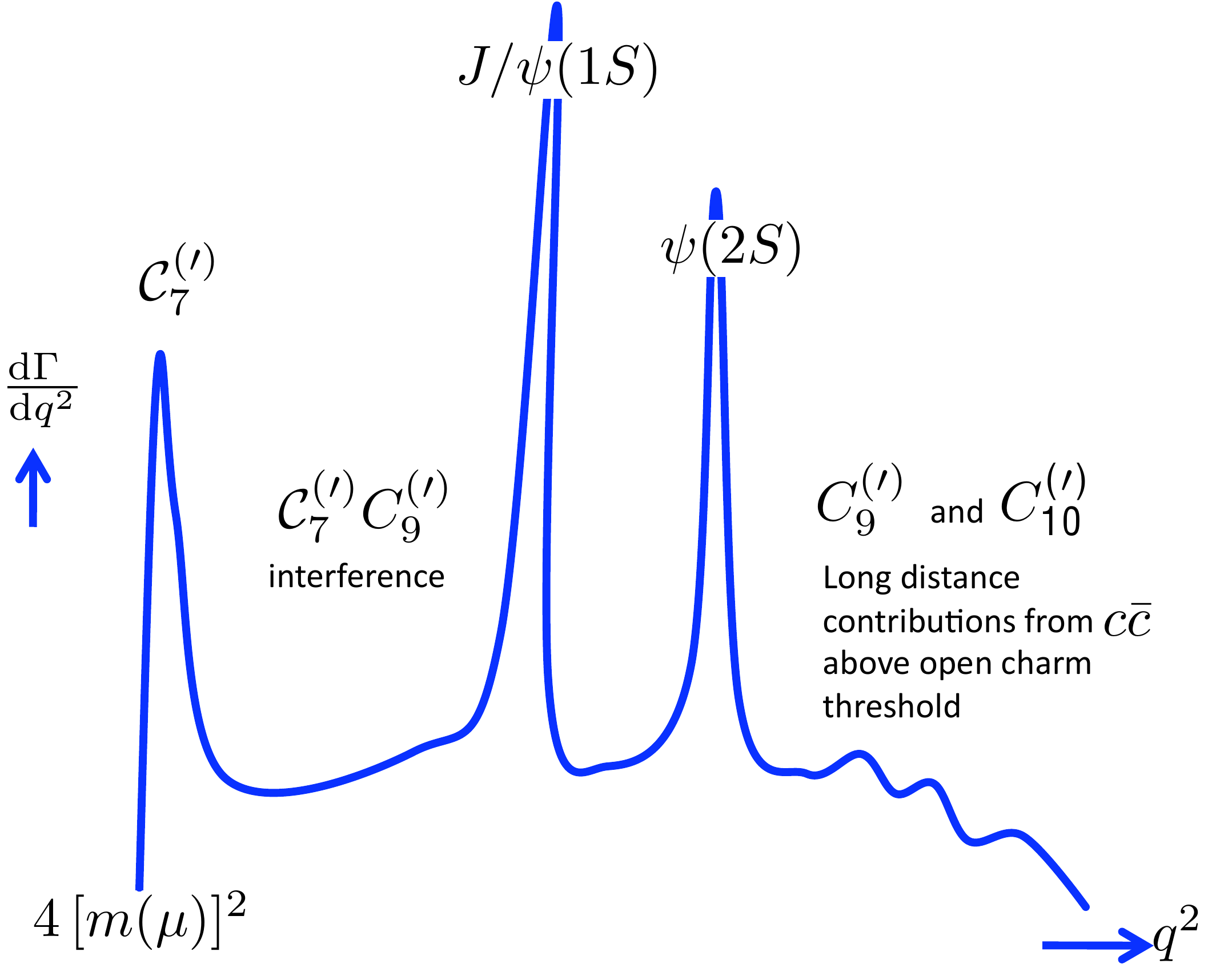}
\end{subfigure}
\caption{(left) Electroweak box and penguin diagrams describing the transition of a \bquark quark into an \squark quark and two leptons. (right) Sketch of the \BdToKstmm differential branching fraction as a function of the dilepton invariant mass squared. Different Wilson coefficients can be probed in different \qsq regions.}
\label{fig:TH}
\end{figure}

Flavour Changing Neutral Current (FCNC) processes, where a quark changes its flavour without altering its electric charge, are forbidden at tree level in the Standard Model (SM) and can only occur via loop diagrams.
This makes such transitions rare and, due to the lack of a dominant tree-level SM contribution, sensitive to new unobserved particles that can show up either as a sizeable increase or decrease in the rate of particular decays, or as a change in the angular distribution of the particles in the detector.
A good laboratory to study FCNC are decays of a \bquark quark into an \squark quark and two leptons, \bTosll, which are described by the electroweak diagrams shown in Fig.~\ref{fig:TH} (left).

The \lhcb detector~\cite{Alves:2008zz,LHCb-DP-2014-002} is a single arm spectrometer fully instrumented in the forward region and designed to study heavy-flavoured hadrons.
During Run-1, \lhcb collected about 1 and 2~\invfb of \pp collisions at centre-of-mass energies of 7 and 8~\tev, respectively.
Due to the large production cross-section in the forward direction these data provide unprecedentedly large numbers of \B and \Lb hadron decays.
A flexible trigger system, excellent momentum and impact parameter resolutions, and the most performant vertexing and particle identification capabilities at the \lhc, make \lhcb the ideal place to look for New Physics (NP) through precise studies of rare \bquark-quark processes.
Recent measurements of semileptonic \bquark-hadron decays are discussed.

%%%%%%%%%%%%%%%%%%%%%%%%%%%%%%%%%%%%
\section{Branching fractions}

\begin{figure}[h!]
\vspace{0.5cm}
\centering
\includegraphics[width=0.36\textwidth]{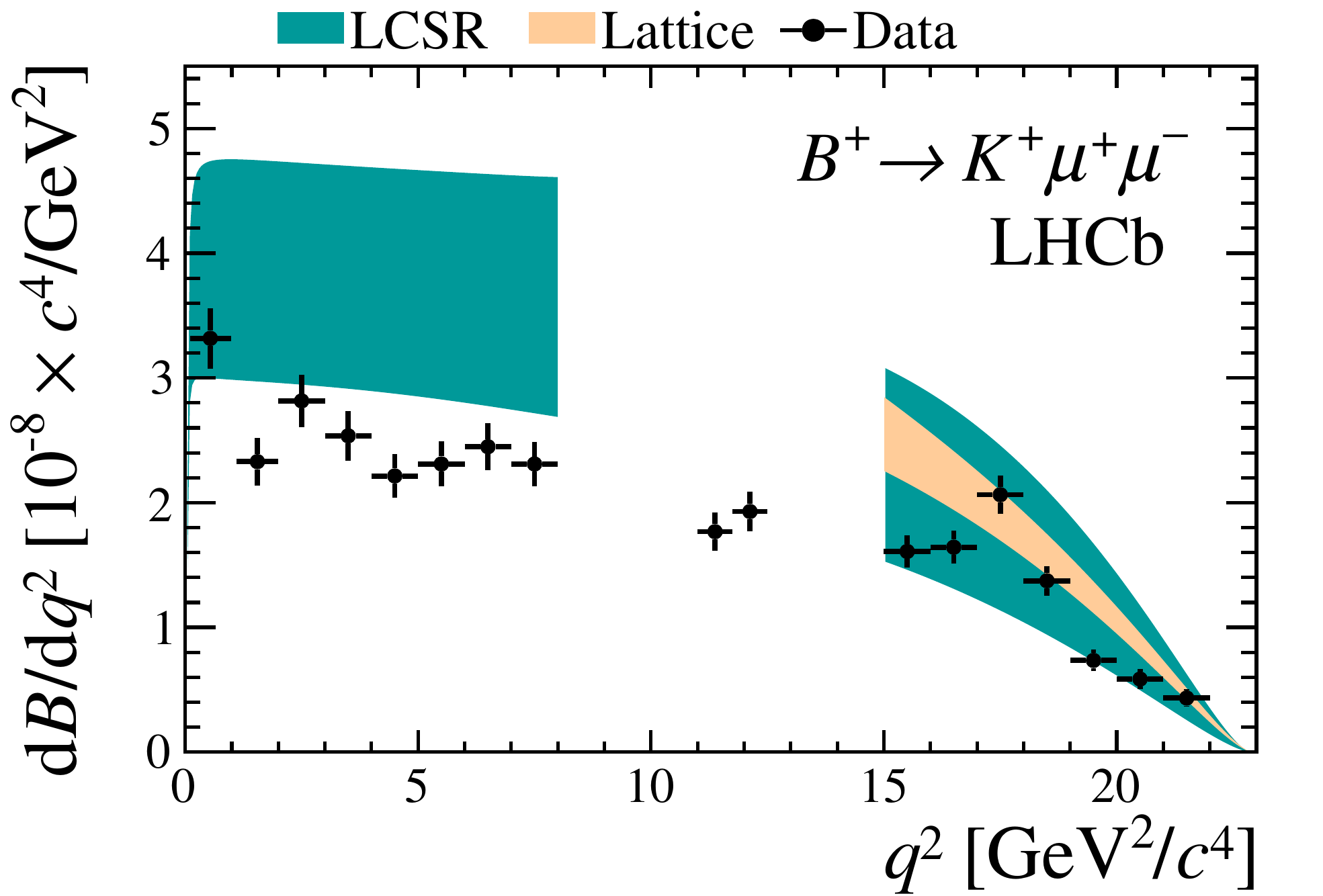}
\includegraphics[width=0.36\textwidth]{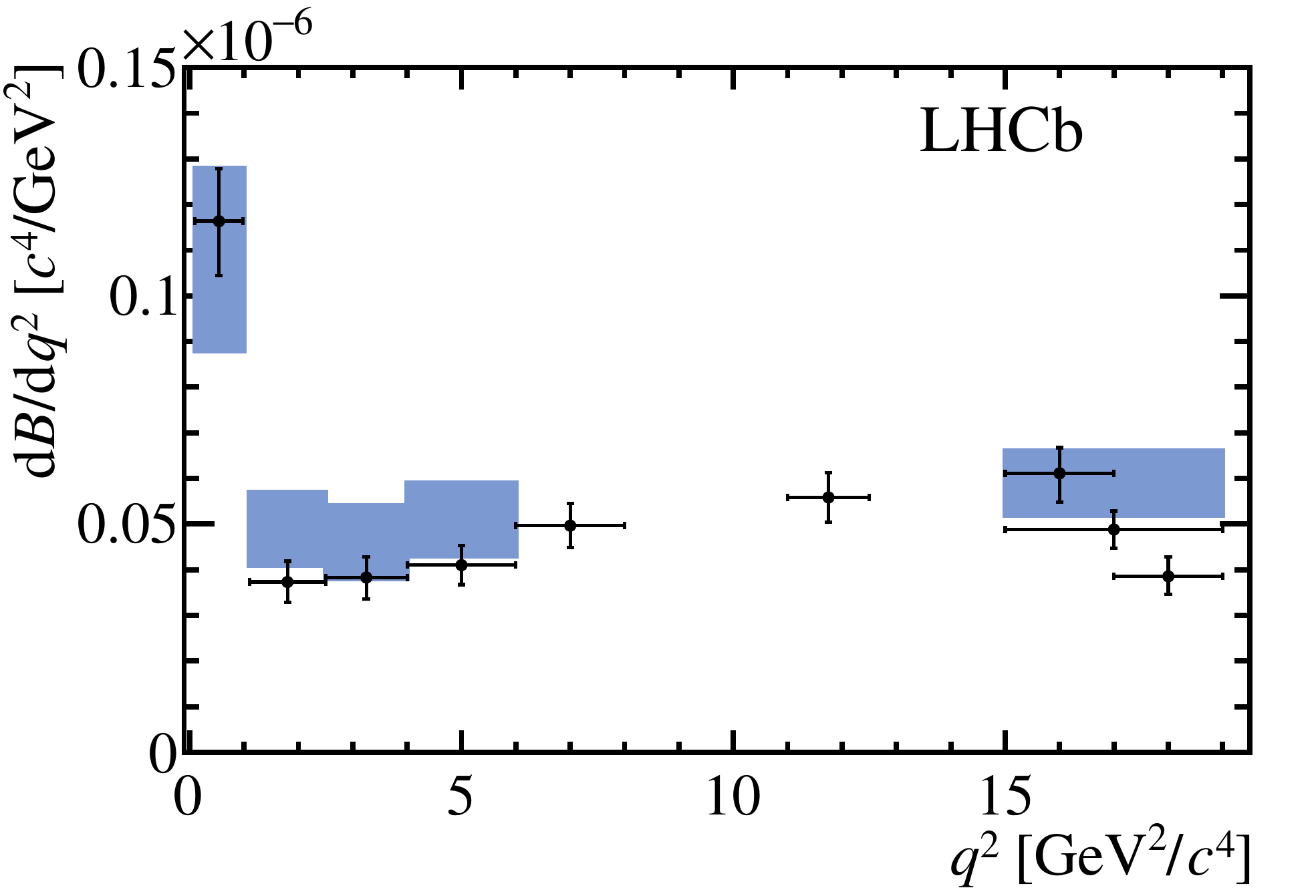}
\includegraphics[width=0.36\textwidth]{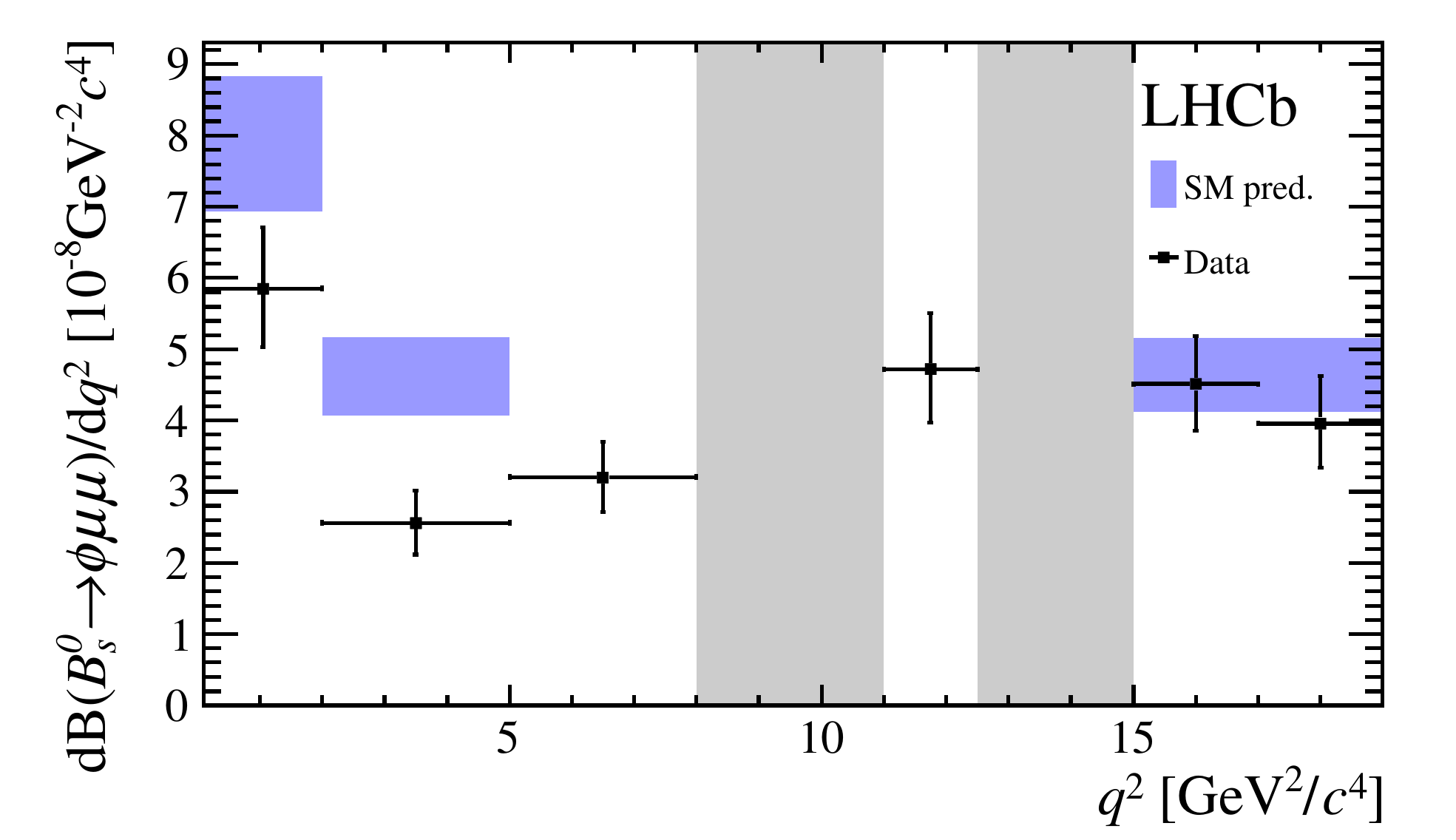} \\
\includegraphics[width=0.36\textwidth]{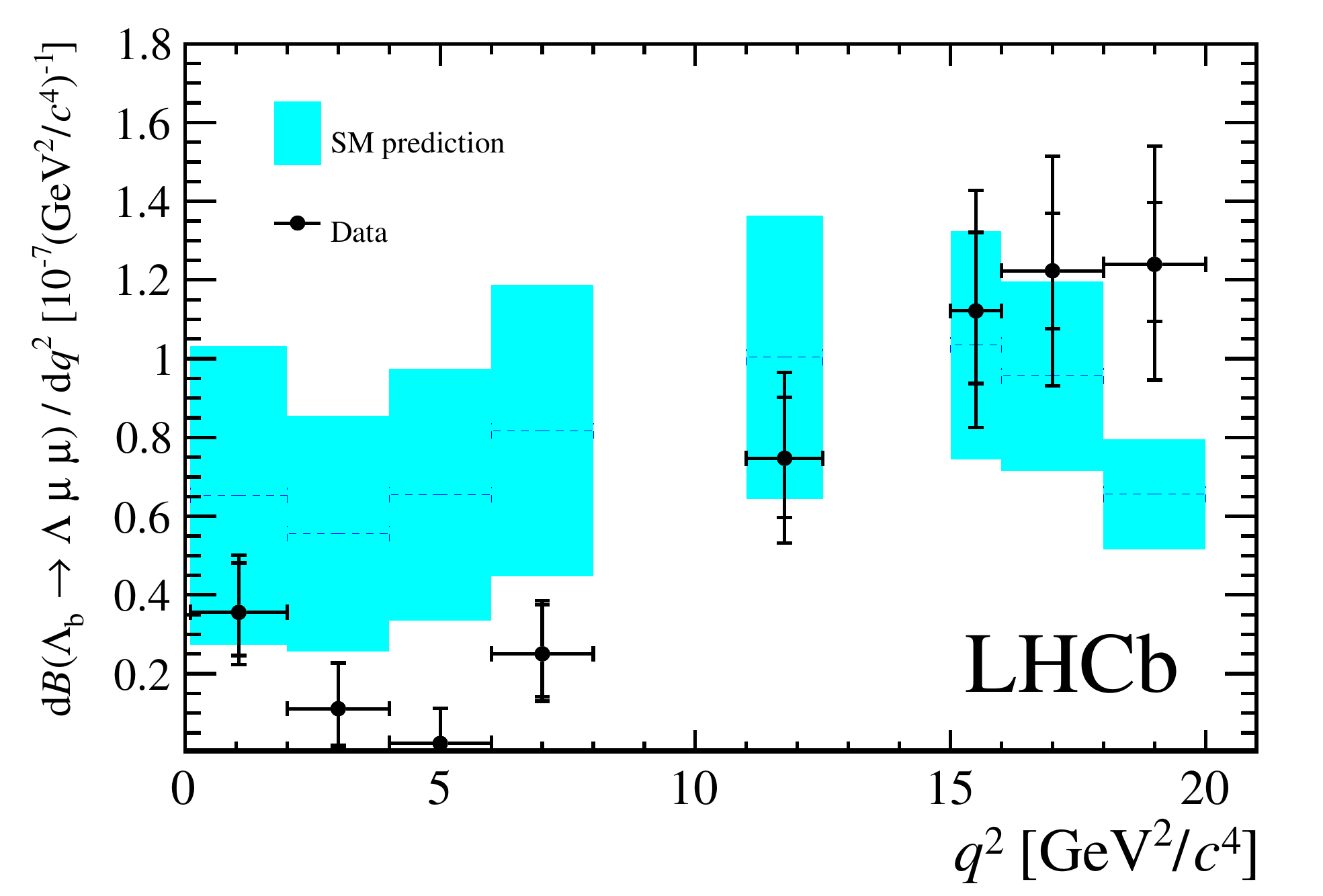}
\includegraphics[width=0.36\textwidth]{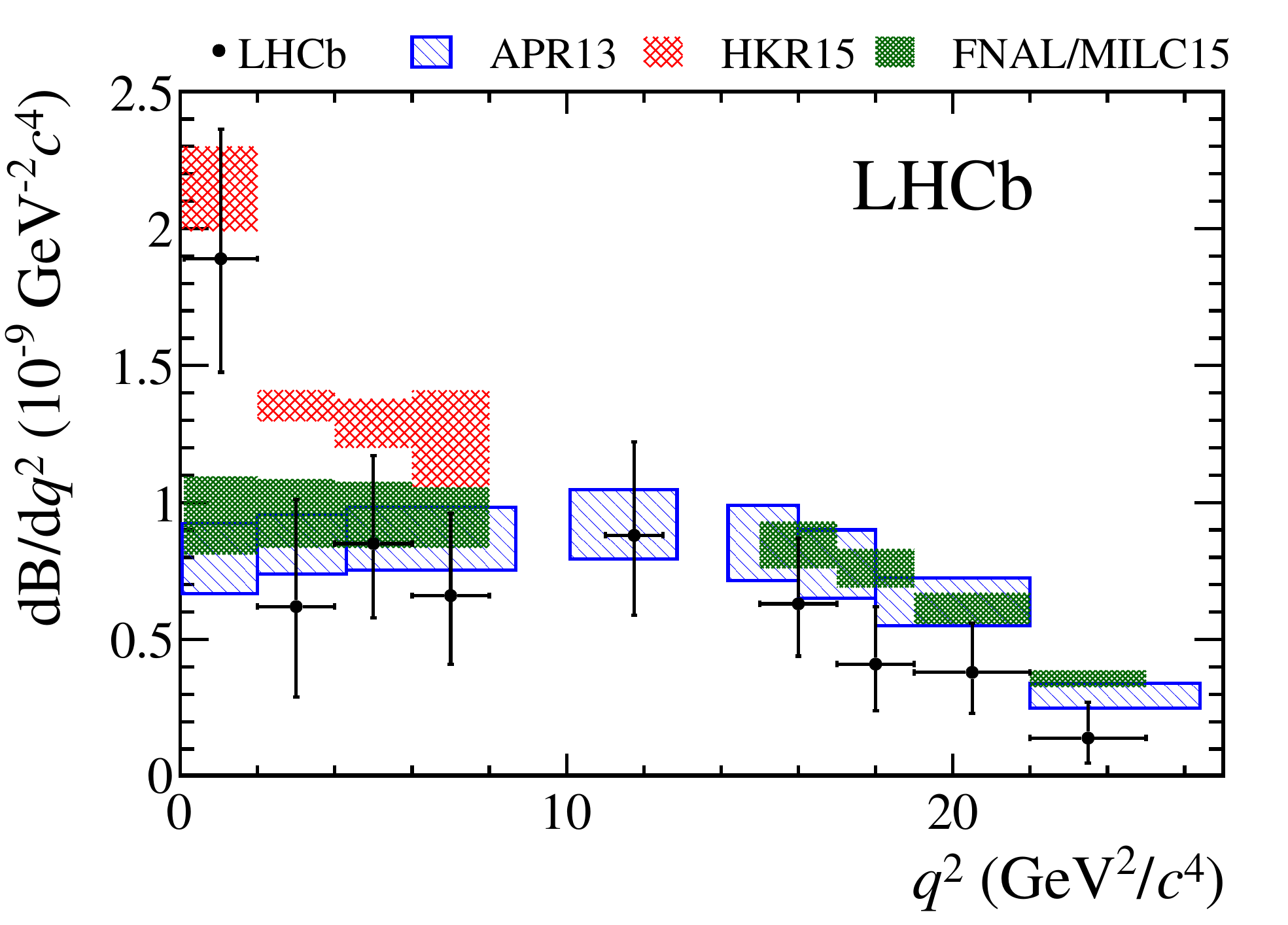}
\caption{Differential branching fraction of the decay (top) \BuToKmm~\protect\cite{LHCb-PAPER-2014-006}, \BdToKstmm~\protect\cite{LHCb-PAPER-2016-012}, \mbox{\BsToPhimm}~\protect\cite{LHCb-PAPER-2015-023} and (bottom) \LbToLmm~\protect\cite{LHCb-PAPER-2015-009}, \BuToPimm~\protect\cite{LHCb-PAPER-2015-035}, compared to SM predictions.}
\label{fig:BF}
\end{figure}

The most basic observable that physics beyond the SM can affect is the rate at which  a particular decay occurs, which motivated the \lhcb collaboration to perform the measurement of the branching fraction of a series of FCNC processes.
All measurements are performed as a function of the dilepton invariant mass squared, \qsq, and compared against the SM prediction.
Final states with muons are considered as being experimentally easier to reconstruct.
The dependence of the \BdToKstmm differential branching fraction as a function of \qsq is shown in Fig.~\ref{fig:TH} (right).
Figure~\ref{fig:BF} shows the differential branching fraction of a collection of \bTosll decays of \B and \Lb hadrons, as well as partner transitions such as \bTodll.
In all cases, the experimental uncertainty is dominated by the limited statistics of the samples available in the Run-1 data set.

In the region below $\sim 6~\gevgevcccc$ in \qsq, the SM predictions consistently overshoot the data, a common trend that is observed both in the mesonic and in the baryonic sectors.
The largest deviation is found in the decay \BsToPhimm in the region $1 < \qsq < 6~\gevgevcccc$, where the data are $3.3\sigma$ below the prediction~\cite{LHCb-PAPER-2015-023}.
Decays of \Lb hadrons are also overestimated in the SM, however predictions are currently much less precise than for \B mesons~\cite{LHCb-PAPER-2015-009}.
Finally, although the \mbox{\BuToPimm} branching fraction is generally compatible with the prediction, agreement in the lowest-\qsq bin is only achieved when contributions from $\rho$ and $\omega$ resonances are taken into account~\cite{LHCb-PAPER-2015-035}.

%%%%%%%%%%%%%%%%%%%%%%%%%%%%%%%%%%%%
\section{Angular analyses}

Besides using branching fractions, much stronger constraints to possible extensions of the SM can be set by studying the angular distribution of the final state particles of FCNC decays.
Depending on the decay mode and on the size of the available data sample, full or simplified angular analyses have been performed.

\begin{figure}[h!]
\vspace{0.5cm}
\centering
\includegraphics[width=0.36\textwidth]{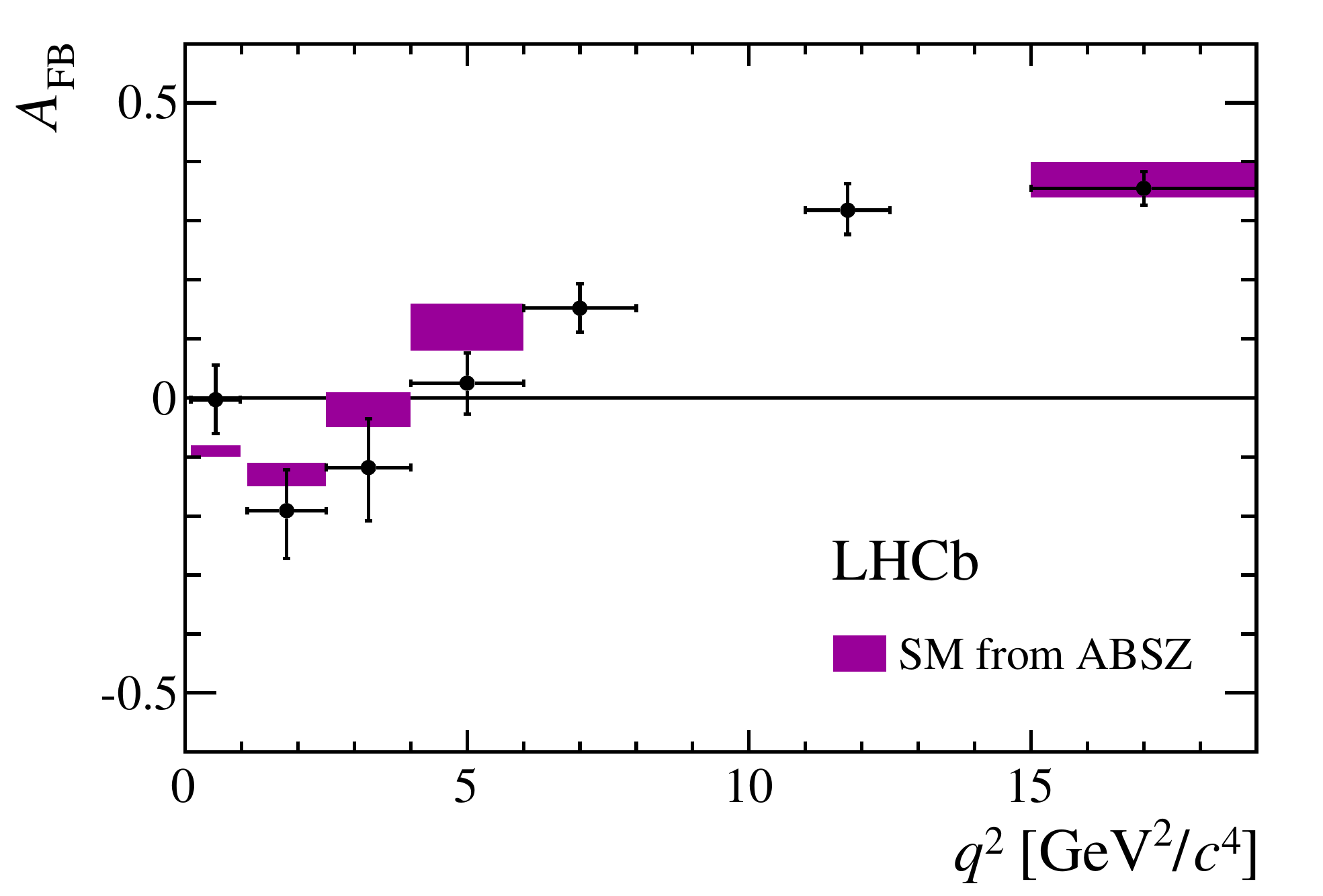}
\includegraphics[width=0.36\textwidth]{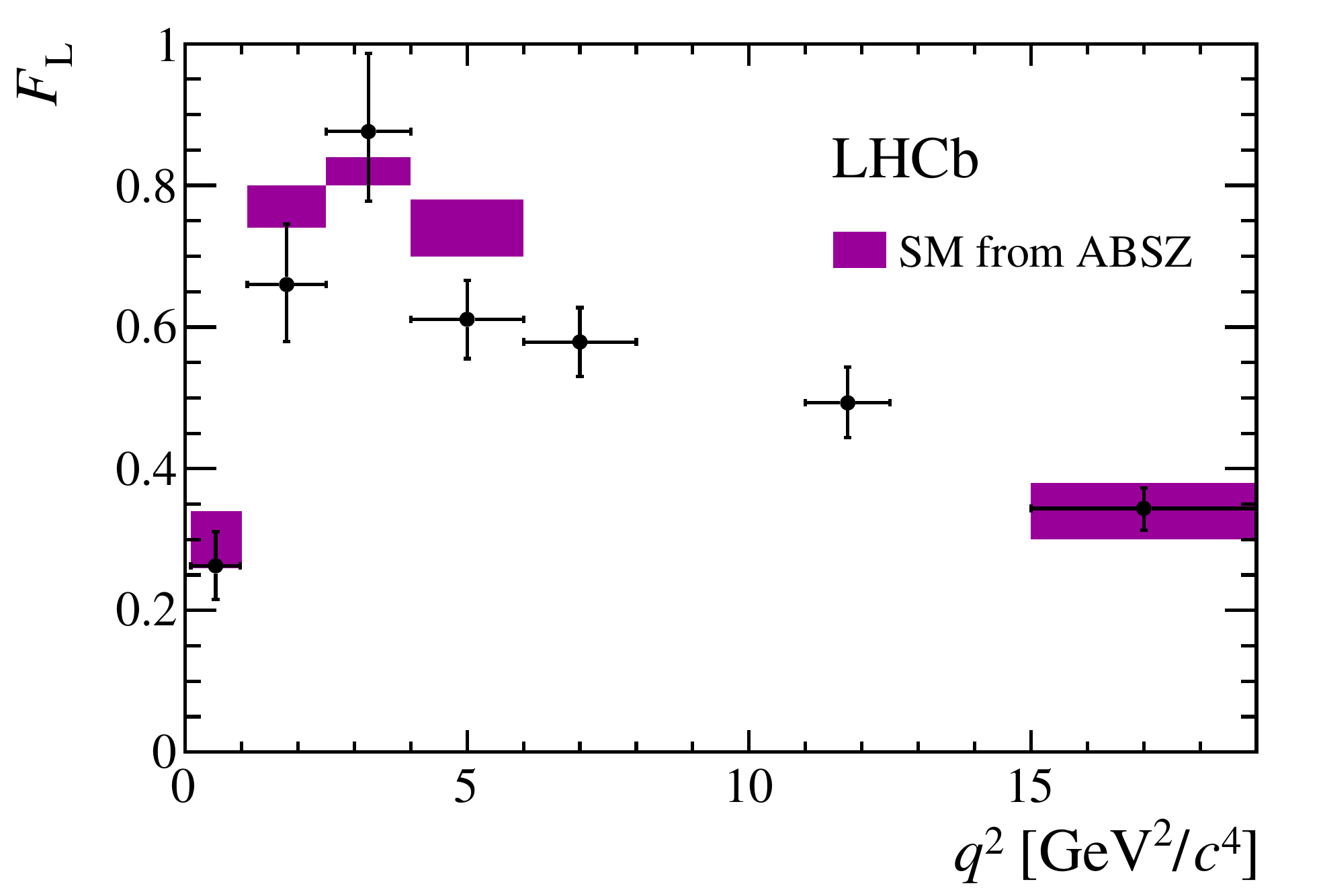}
\includegraphics[width=0.36\textwidth]{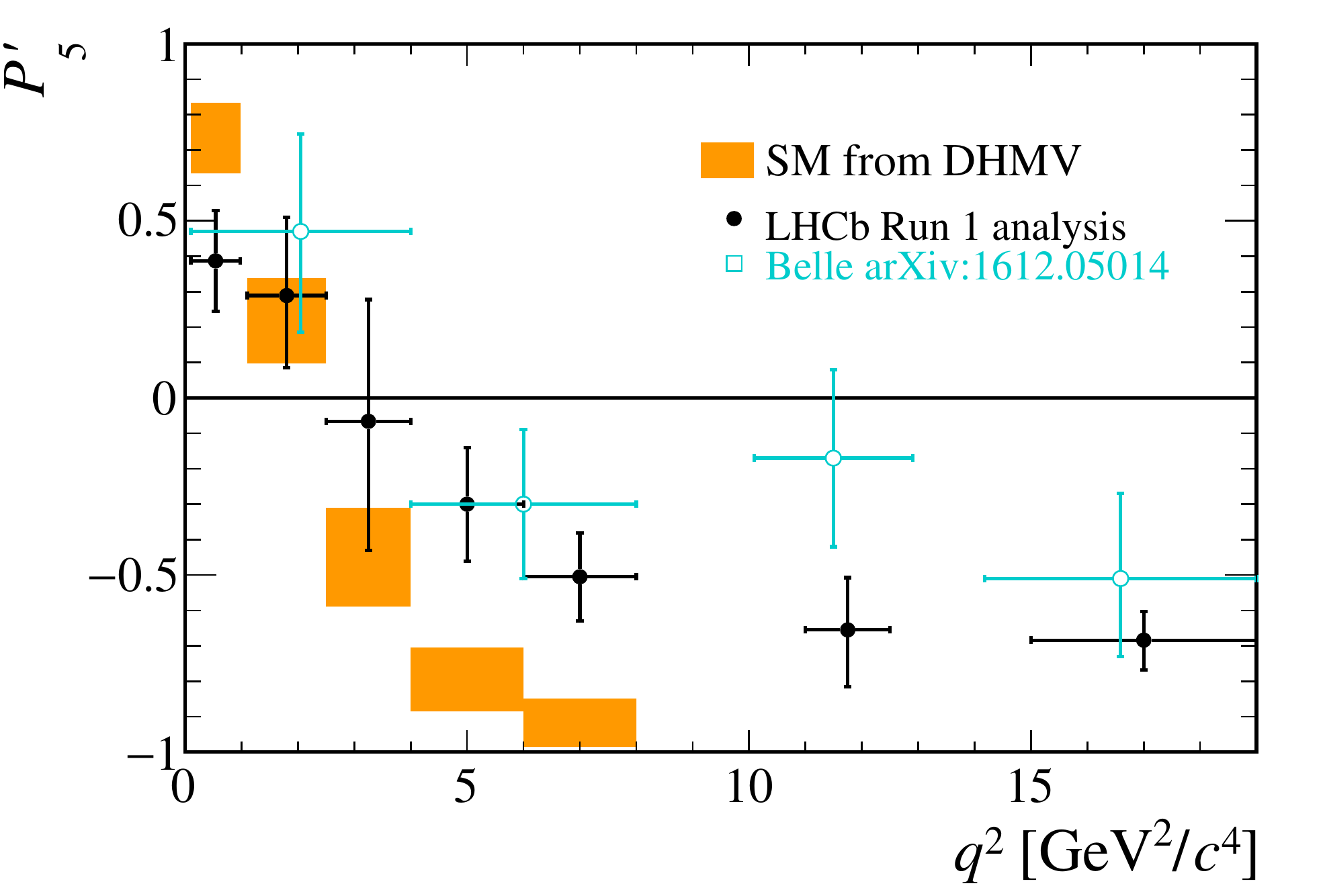} \\
\includegraphics[width=0.36\textwidth]{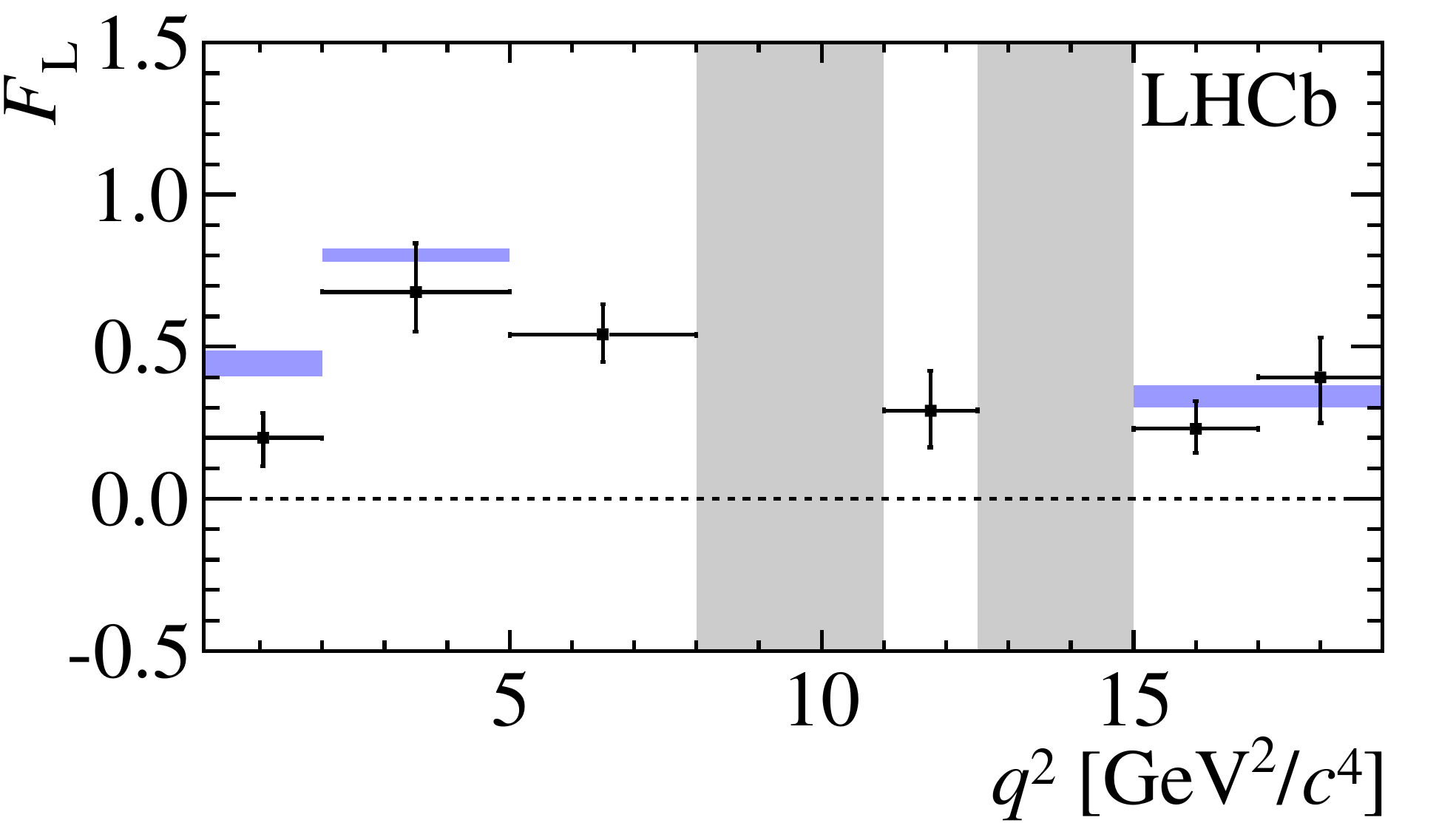}
\includegraphics[width=0.36\textwidth]{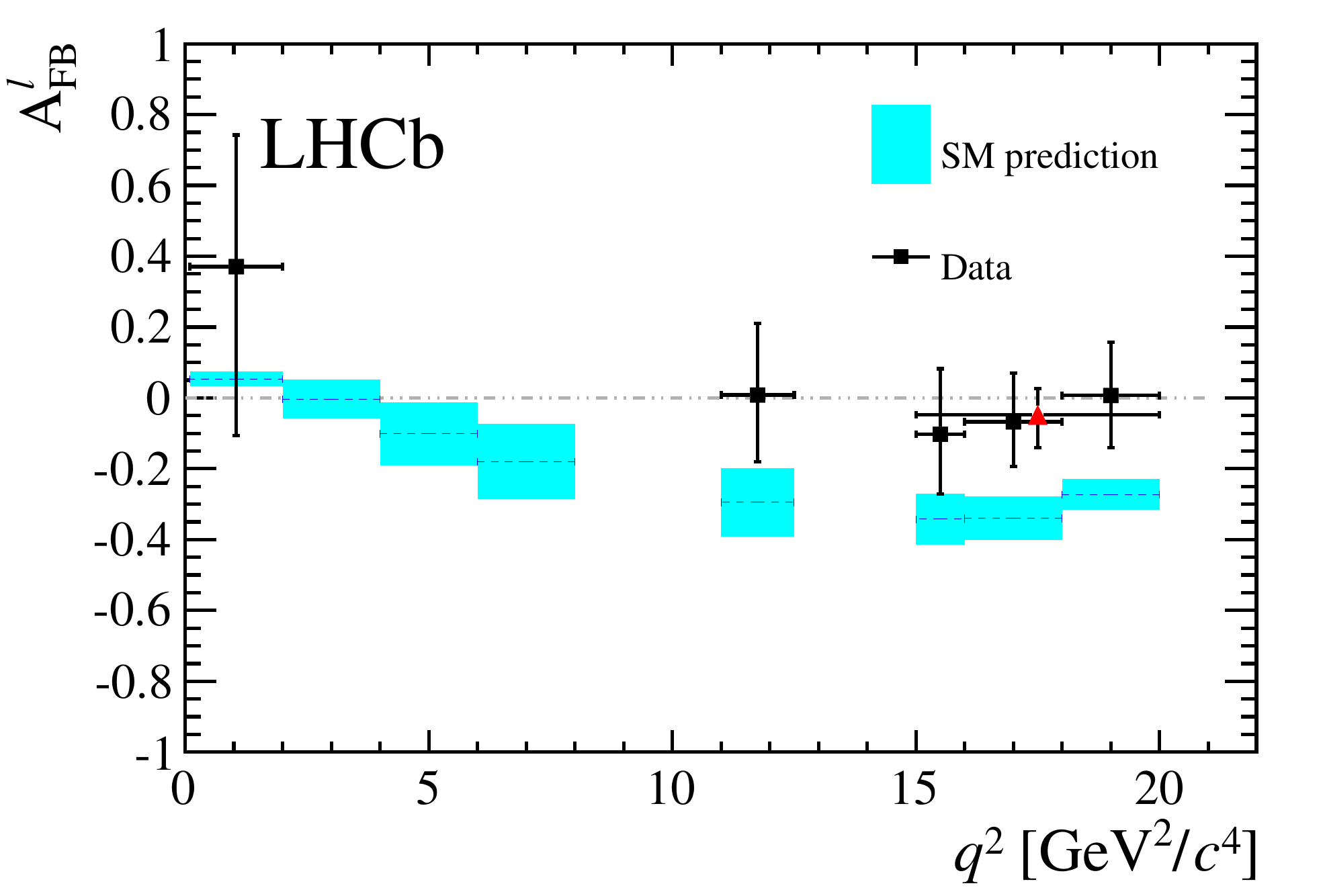}
\caption{Angular observables of the decay (top) \BdToKstmm~\protect\cite{LHCb-PAPER-2016-012} and (bottom) \BsToPhimm~\protect\cite{LHCb-PAPER-2015-023}, \LbToLmm~\protect\cite{LHCb-PAPER-2015-009}, compared to SM predictions.}
\label{fig:AA}
\end{figure}

The decay \BdToKstmm has a complex angular structure that can be fully described by three angles and \qsq, and that provides many observables sensitive to different types of NP.
The \lhcb collaboration has performed the first full angular analysis of this mode, and measured the full set of CP-averaged angular terms, their correlations, as well as the full set of CP-asymmetries~\cite{LHCb-PAPER-2016-012}.
The forward-backward asymmetry of the dimuon system, \afb, and the longitudinal polarisation fraction of the \Kstarz, \fl, compared to the SM prediction are presented in Fig.~\ref{fig:AA} (top).
There is general consistency, but, similarly to the branching fraction, large uncertainties due to the hadronic-matrix elements affect the predictions.
However, it is possible to construct ratios of observables that are less dependent on the form-factors and that can be theoretically determined with improved precision~\cite{Descotes-Genon:2013vna}.
Figure~\ref{fig:AA} (top right) shows one of such observables, \pfp, which manifests a local deviation from the SM in the region between 4 and 8~\gevgevcccc in \qsq of about $3\sigma$.
An angular analysis of the decay \BdToKstee in the \qsq range between 0.002 and 1.120~\gevgevcccc has also been performed by \lhcb~\cite{LHCb-PAPER-2014-066}, where all the measured observables are found to be consistent with the predictions.

Although the decay \BsToPhimm has a reduced number of angular observables that can be accessed compared to \BdToKstll modes, a full angular analysis has also been performed~\cite{LHCb-PAPER-2015-023}.
Figure ~\ref{fig:AA} (bottom left) shows \fl, which is found to be in good agreement with the SM prediction.
Finally, because baryonic transitions allow to extract complementary information to that available in decays of \B mesons, \lhcb has performed the first angular analysis of \LbToLmm~\cite{LHCb-PAPER-2015-009}.
Two forward-backward asymmetries, one in the hadronic and one in the leptonic system, have been measured.
While the former is in good agreement with the SM, the latter is consistently above the prediction, as reported in Fig.~\ref{fig:AA} (bottom right).

%%%%%%%%%%%%%%%%%%%%%%%%%%%%%%%%%%%%
\section{Tests of lepton universality}

Due to the equality of the electroweak couplings of gauge bosons to leptons, 
the decay rate of processes whose final states only differ by the flavour of the participating leptons are expected to be the same in the SM, except from very small Higgs penguin contributions and difference in phase space due to the lepton masses.
In particular, ratios of branching fractions represent a powerful null test of the SM as theoretical uncertainties largely cancel in the predictions, and experimental systematics are much reduced.

\begin{figure}[h!]
\vspace{0.5cm}
\centering
\includegraphics[width=0.45\textwidth]{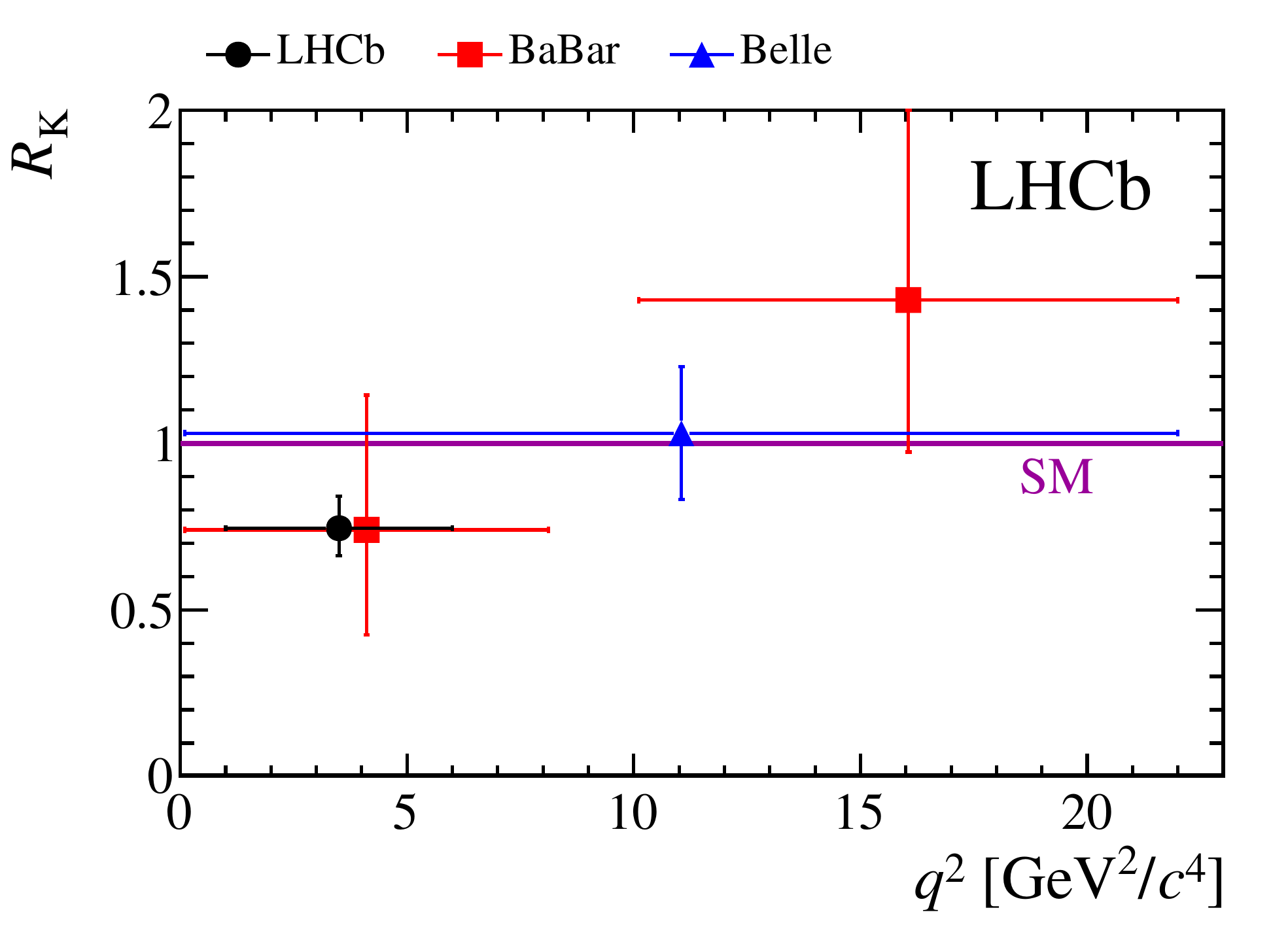}
\caption{Measurement of $\RK = \frac{\BF(\BuToKmm)}{\BF(\BuToKee)}$~\protect\cite{LHCb-PAPER-2014-024}, compared to previous experiments and to the SM prediction.}
\label{fig:RK}
\end{figure}

In the SM the ratio \mbox{$\RK = \frac{\BF(\BuToKmm)}{\BF(\BuToKee)}$} is precisely predicted to be \mbox{$1 \pm \mathcal{O}(10^{-3})$}~\cite{Hiller:2003js}.
The \lhcb collaboration has performed the most precise test of lepton universality using \BuToKmm and \BuToKee decays to date~\cite{LHCb-PAPER-2014-024}.
The experimental result in the range $1 < \qsq < 6~\gevgevcccc$ is $\RK = 0.745^{+0.090}_{-0.074}\stat \pm 0.036\syst$, which manifests a tension with the SM prediction at the level of $2.6\sigma$, as shown in Fig.~\ref{fig:RK}.

\begin{figure}[h!]
\vspace{0.5cm}
\centering
\includegraphics[width=0.45\textwidth]{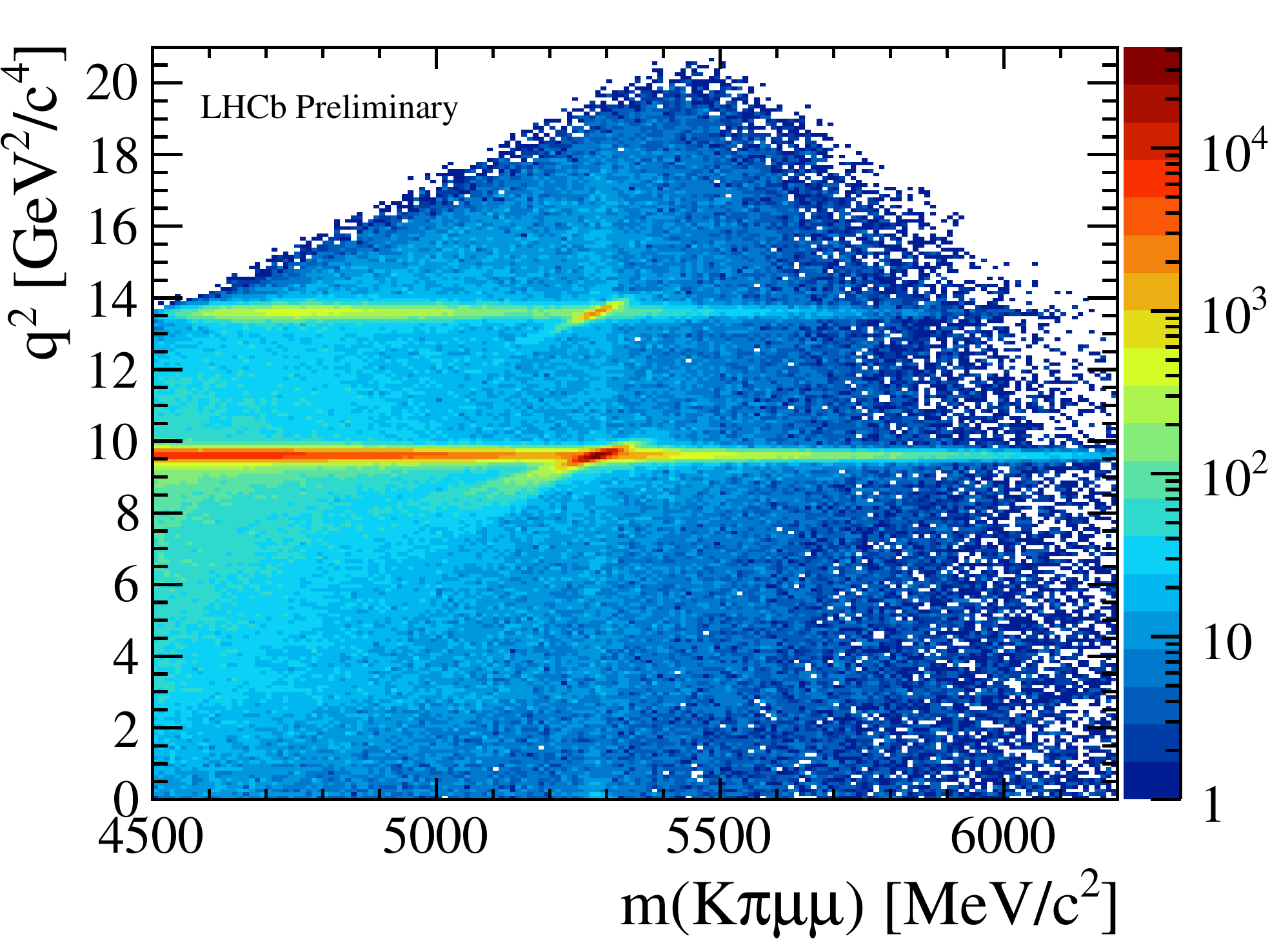}
\hspace{1cm}
\includegraphics[width=0.45\textwidth]{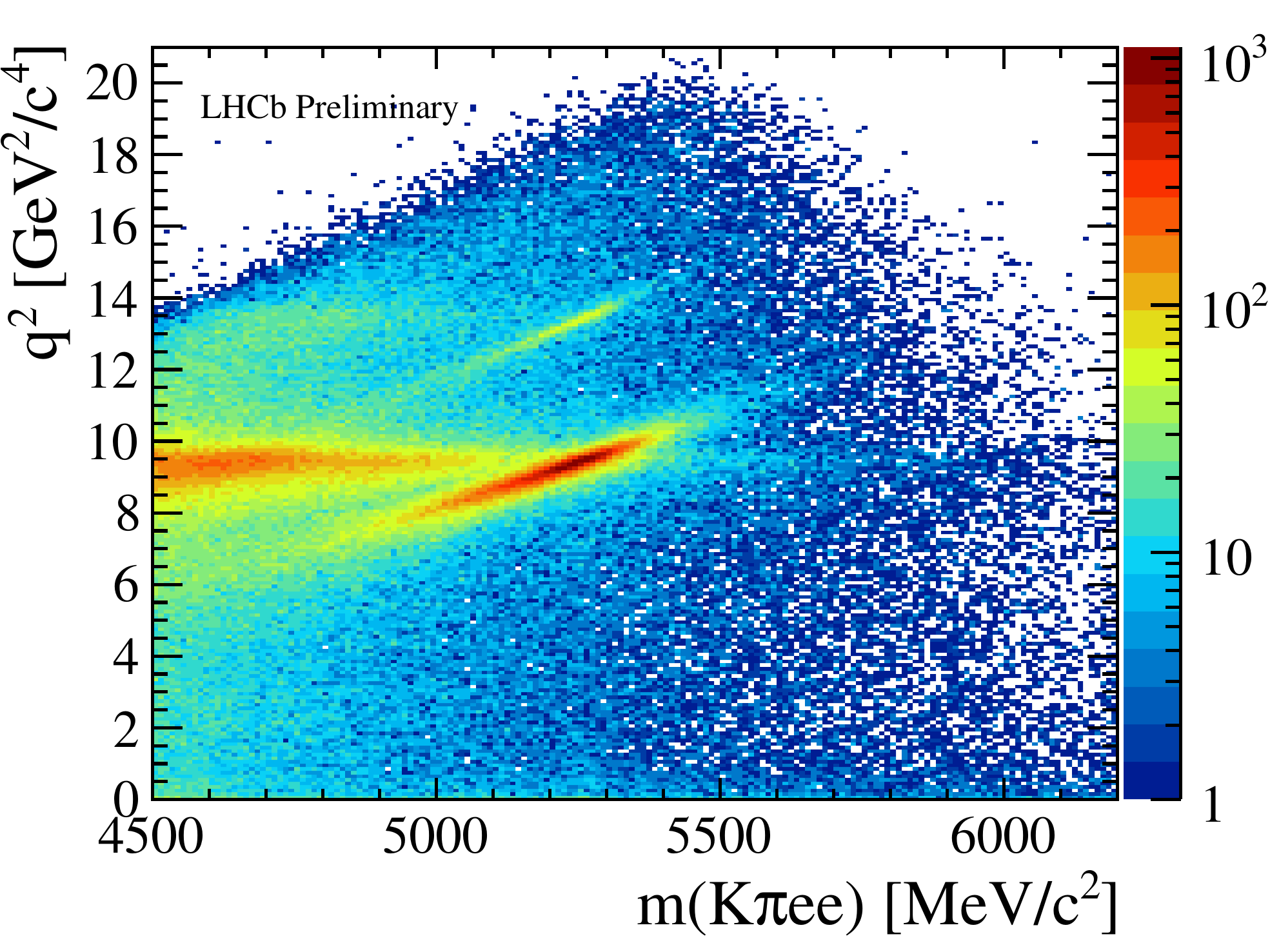}
\caption{Di-lepton invariant mass squared, \qsq, as a function of the four-body invariant mass of the \Bz candidates for \BdToKstll final states with (left) muons and (right) electrons.}
\label{fig:RKst}
\end{figure}

Several other tests of lepton universality are currently being carried out by the \lhcb collaboration, most notably the measurement of the ratio of the branching fractions of \BdToKstmm and \BdToKstee, \RKst.
The experimental environment in which the detector operates leads to significant differences in the treatment of decays involving muons or electrons in the final state.
Particularly, the much larger amount of bremsstrahlung emitted by the electrons and the different trigger thresholds due to the higher occupancy of the calorimeters than of the muon stations result in a difference in the reconstruction efficiency of about a factor five in favour of the muonic mode.
Figure~\ref{fig:RKst} shows the distribution of the di-lepton invariant mass squared as a function of the four-body invariant mass of the \Bz candidates for final states with muons and electrons, where the second exhibit a more prominent radiative tail of the \jpsi and \psitwos resonances as well as a stronger contamination by partially-reconstructed backgrounds.
The efficiency corrections are controlled by measuring the ratio $\frac{\BF(\BdToKstJPsmm)}{\BF(\BdToKstJPsee)}$, as well as determining \RKst as a double ratio to the resonant modes \BdToKstJPsll.
A preliminary result has been presented at a CERN-LHC Seminar~\cite{seminar}.

Independent tests of lepton universality are also being carried out using \decay{\bquark}{\cquark\ell\nu_{\ell}} decays with tau leptons in the final state.
The ratio of branching fractions $\RDst = \frac{\BF(\decay{\Bdb}{\Dstarp\taum\neutb})}{\BF(\decay{\Bdb}{\Dstarp\mun\neumb})}$ is measured to be $\RDst = 0.336 \pm 0.027\stat \pm 0.030\syst$, which is $2.1\sigma$ larger than the value expected in the SM~\cite{LHCB-PAPER-2015-025}.

%%%%%%%%%%%%%%%%%%%%%%%%%%%%%%%%%%%%
\section{Summary and conclusions}

Recent rare decays of \B and \Lb hadrons performed by the \lhcb collaboration have been presented.
While most of the observables are in good agreement with the SM predictions, some intriguing tensions have been observed, most notably in branching fractions of \bTosll processes in the low region of \qsq, in the \pfp angular observable in \BdToKstmm, and in the \RK and \RDst ratios.

Several attempts to interpret these anomalies have been made by performing global fits to various $\bquark \to \squark$ observables from different experiments~\cite{Altmannshofer:2015sma, Descotes-Genon:2015uva,Ciuchini:2015qxb}.
All these fits point to a tension between the data and the SM with a significance of around 3--$4\sigma$.
Different models have been proposed to account for such effects, for example containing a new heavy gauge boson $Z'$~\cite{Descotes-Genon:2013wba,Gauld,Buras,Altmannshofer,Altmannshofer:2014cfa,Altmannshofer:2016jzy} or leptoquarks~\cite{Becirevic:2016yqi,Crivellin:2017zlb}, but a definitive explanation has yet to be found.

The current status strongly motivates further work both in the theory and in the experimental side to clarify the present observations.
With the upcoming Run-2 data \lhcb will continue to perform analyses of rare \bquark-quark decays, including additional tests of lepton universality such as \RPhi.

%%%%%%%%%%%%%%%%%%%%%%%%%%%%%%%%%%%%
\section*{References}
\bibliography{my-bibliography,LHCb-PAPER,LHCb-DP}
%%%%%%%%%%%%%%%%%%%%%%%%%%%%%%%%%%%%
\end{document}